\documentclass[preprint,english,aps,showpacs]{revtex4-1}
\pdfoutput=1
\usepackage{amsmath}
\usepackage{color}
\usepackage{graphicx}
\usepackage{epstopdf}

\usepackage{babel}

\begin{document}

\title{Real structure of lattice matched GaAs-Fe$_3$Si core-shell nanowires}
\author{B.~Jenichen}
\email{bernd.jenichen@pdi-berlin.de}
\author{M.~Hilse}
\author{J.~Herfort}
\author{A.~Trampert}
\address{Paul-Drude-Institut f\"ur Festk\"orperelektronik,
Hausvogteiplatz 5--7,D-10117 Berlin, Germany}

\date{\today}

\begin{abstract}
GaAs nanowires and GaAs-Fe$_{3}$Si core-shell nanowire structures were grown by molecular-beam epitaxy on oxidized Si(111) substrates and characterized by transmission electron microscopy (TEM) and X-ray diffraction (XRD). Ga droplets were formed on the oxide surface, and the semiconducting GaAs nanowires grew epitaxially via the vapor-liquid-solid mechanism as single-crystals from holes in the oxide film. We observed two stages of growth of the GaAs nanowires, first the regular growth and second the residual growth after the Ga supply was finished. The magnetic Fe$_{3}$Si shells were deposited in an As-free chamber. They completely cover the  GaAs cores although they consist of small grains. High-resolution TEM micrographs depict the differently oriented grains in the Fe$_3$Si shells. Selected area diffraction of electrons and XRD gave further evidence that the shells are textured and not single crystals. Facetting of the shells was observed, which lead to thickness inhomogeneities of the shells.
\end{abstract}

%%\pacs{68.70.+w, 68.55.ag, 68.37.Lp, 61.05.cp}
%%61.05.cp xrd, 68.37.Lp TEM, 68.55.ag semiconductors,68.70.+w, whiskers

%%\keywords{A1 Nanostructures; A3 Molecular beam epitaxy; B1 Gallium compounds; B1 Metals; B2 Magnetic materials}

\maketitle

\section{Introduction}

Semiconductor nanowires (NWs) represent systems for exploring nanoscale physics and to design a variety of new devices (see Ref.~\cite{lu2006} for a review). They can be grown not only on dissimilar substrates but also as axial and radial heterostructures.\cite{Lieber2007,couto12,Breuer2011}
Device concepts based on the spin rather than the charge of the electron have been introduced in the field of spintronics. Among these concepts, nanowires that combine a semiconductor and a ferromagnet in a core-shell geometry have gained a lot of interest since 2009 when they were presented for the first time \cite{Hilse2009,Rudolph2009,Rueffer2012,Dellas2010,Tivakorn2012,Yu2013}. Because of the cylindrical shape of the ferromagnet, such core-shell nanowires could allow for a magnetization along the wire and thus perpendicular to the substrate surface. Ferromagnetic stripes or tubes with a magnetization perpendicular to the substrate have the potential for circular polarized light emitting diodes that optically can transmit spin information in zero external magnetic field and thus allow for on-chip optical communication of spins on the one hand \cite{Farshchi2011}. On the other hand they enable three-dimensional magnetic recording with unsurpassed data storage capacities \cite{Parkin2008,Ryu2012}.
The perfect lattice matching of the binary Heusler alloy Fe$_{3}$Si and GaAs allows for the molecular beam epitaxy (MBE) growth of planar high quality hybrid structures \cite{Jenichen05,Herfort2006,Herfort2006b}. In addition, the cubic Fe$_{3}$Si phase \cite{Hansen1958,Elliot1965}, shows a robust stability against stoichiometric variations with only slightly modified magnetic properties \cite{Herfort2004}. Moreover, its thermal stability against chemical reactions at the ferromagnet/semiconductor interface is considerably higher than that of conventional ferromagnets like Fe, Co, Ni, and Fe$_{x}$Co$_{1-x}$ \cite{herfort03}.  Together with the high Curie temperature of about 840 K this material system has therefore several advantages compared to most of the previously studied semiconductor-ferromagnet core-shell nanowires using ferromagnetic materials that cannot reach the high quality of a binary Heusler alloy like Fe$_{3}$Si \cite{Hilse2009,Rudolph2009,Rueffer2012,Dellas2010,Tivakorn2012,Yu2013}. Recently, we have demonstrated for the first time that GaAs-Fe$_{3}$Si core-shell NWs prepared by MBE show ferromagnetic properties with a magnetization oriented along the NW axis (perpendicular to the substrate) \cite{hilse2013}. However, the structural properties and hence the magnetic properties of the core-shell NWs depend strongly on the substrate temperature during the growth of the Fe$_{3}$Si shell \cite{hilse2013}. In this work, we present a detailed investigation of the real structure of both GaAs  NWs and GaAs-Fe$_{3}$Si core-shell NWs using transmission electron microscopy (TEM) and X-ray diffraction (XRD) .

\section{Experiment}

Fe$_{3}$Si-GaAs core-shell NW structures are grown by MBE on Si(111) substrates. First, GaAs nanowires are fabricated by the Ga-assisted growth mode on the Si(111) substrates covered with a thin native Si-oxide layer. The growth mechanism is the vapor-liquid-solid (VLS) mechanism,\cite{Wagner1964,Mandl2006,glas2007,Fontcuberta2008,Wacaser2009} where pin holes in the SiO$_2$ serve as nucleation sites.\cite{Breuer2011}  A Ga droplet is the preferred site for deposition from the vapor. The GaAs NW then starts to grow by preferential nucleation at the spatially restricted GaAs/Si interface (IF). Further growth is unidirectional and proceeds at the solid/liquid IF.  The GaAs NWs are grown at a substrate temperature of 580$^\circ$C, and a V/III flux ratio of unity. The equivalent two-dimensional growth rate amounts to 100~nm/h. In order to finish the NW growth, the Ga-shutter is closed. Then the Ga-droplets on top of the NWs are consumed in the arsenic atmosphere. During this phase of the experiment there is still a certain NW growth taking place at a reduced diameter. The samples are then cooled down.
Once the NW templates are grown, they are transferred under ultra high vacuum conditions to an As free growth chamber for metals of the same MBE system.  There the GaAs NW templates are covered with Fe$_3$Si shells  at different substrate temperatures ranging from 100~$^\circ$C to 350~$^\circ$C. More details regarding the growth conditions can be found in Ref.~\cite{hilse2013}.
 The main growth parameters are summarized in Table \ref{tab:tab1}.

\begin{table*}[htbp]
% see materials.xlsx
  % \centering
  \caption{
   Substrate temperatures T$_S$ during epitaxial NW growth and equivalent film thicknesses, calculated from growth rates and deposition times for the four samples investigated.}

    \begin{tabular}{l c c c c c c c c}
    \hline
          & sample & No.~0 & sample & No.~1 & sample & No.~2 & sample & No.~3 \\
          &    thickness    &    T$_S$    &  thickness    &     T$_S$ &  thickness    &     T$_S$ &  thickness    &     T$_S$\\
    %\midrule

          & (nm)  & $^\circ$C & (nm) & $^\circ$C & (nm) & $^\circ$C & (nm) & $^\circ$C \\
    \hline

    \multicolumn{1}{c}{GaAs} & 22    & 580 & 22 & 580 & 22 & 580 & 22 & 580 \\
    \multicolumn{1}{c}{Fe$_{3}$Si} & -    & - & 69 & 100 & 69 & 200 & 69 & 350 \\
    %%\multicolumn{1}{c}{No.} & m53324    & - & m53325 & m53321 & m53336 & - \\
    %%\multicolumn{1}{c}{Fe$_{3}$Si} & 45 (60)    & 100 & 45 (53-62) & 100 & 9 (9) & 150 \\
    %\bottomrule
        \hline
    \end{tabular}%
  \label{tab:tab1}%
\end{table*}%

The NW structures are characterized by scanning electron microscopy (SEM), by dark-field (DF) and
high-resolution (HR) TEM, selected area diffraction (SAD) of electrons and XRD. The TEM
specimens are prepared by mechanical lapping and polishing,
followed by argon ion milling according to standard techniques.
TEM images are acquired with a JEOL 3010 microscope operating at
200 kV and 300~kV. The cross section TEM methods provide
high lateral and depth resolutions on the nanometer scale, however they average
over the thickness of the thin sample foil or the thickness of the NW as a whole.
The resolution limit of the dark-field method with the TEM used is in the ideal case about 0.2 nm. There can be additional errors due to projection and due to curvature of the interfaces.

HR XRD measurements are performed
using a Panalytical X-Pert PRO MRD\texttrademark\ system
with a Ge(220) hybrid monochromator  (Cu~K$\alpha_1$ radiation with a
wavelength of $\lambda=1.54056$~\AA).

\section{Results and Discussion}

% fig.1
\begin{figure}[!t]
\includegraphics[width=14.0cm]{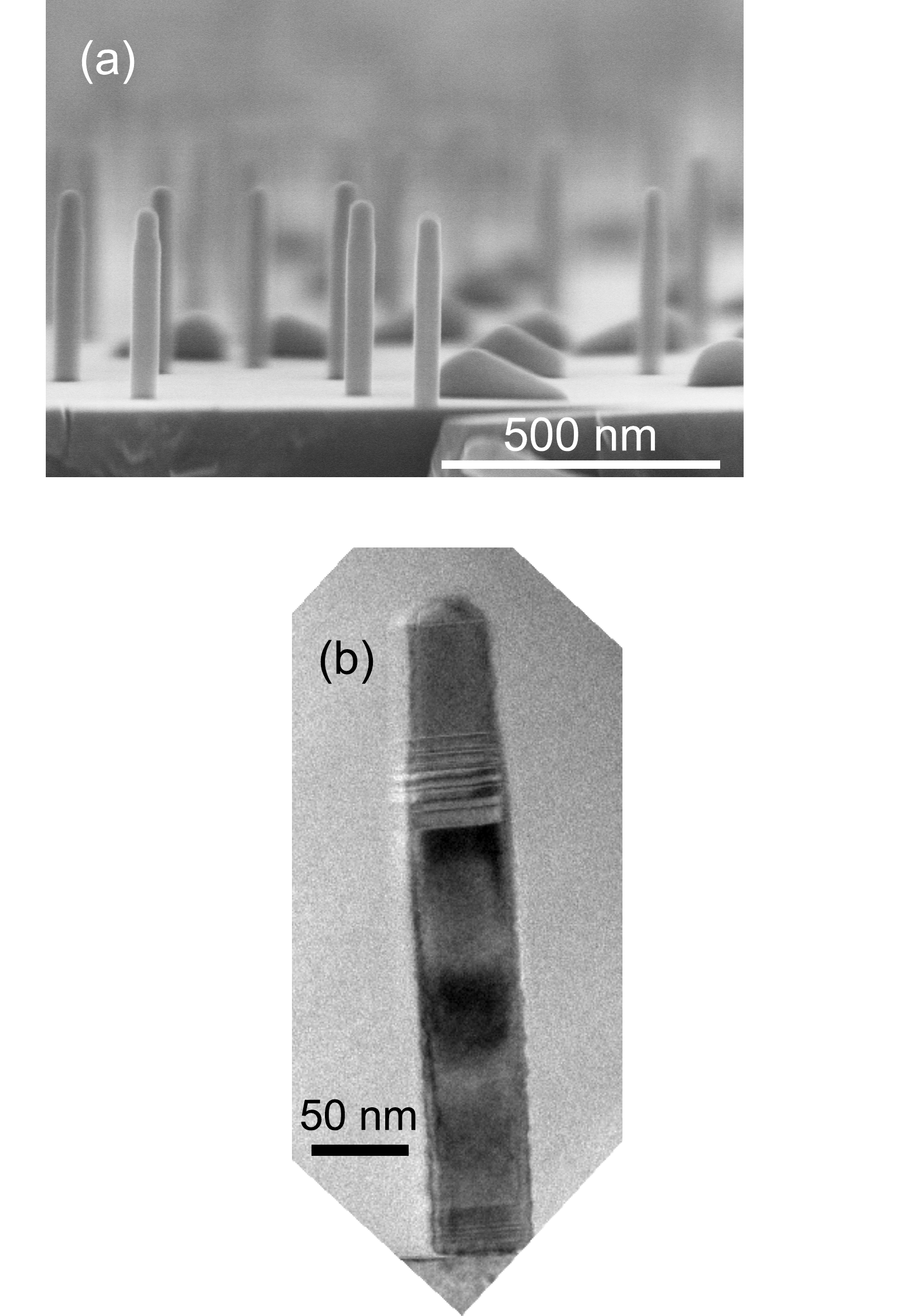}
\caption{(a) SEM image of  GaAs  NWs and GaAs islands between the NWs grown by molecular beam epitaxy on a Si(111) substrate (sample~0). The end pieces of the NWs have a smaller diameter. (b) Multi beam TEM micrograph of a GaAs NW.
}
\label{fig:SEM}
\end{figure}

Figure~\ref{fig:SEM}~(a) shows an SEM micrograph of the pure GaAs NWs (sample~0).  The micrograph of the sample surface reveals a relatively low area density of NWs of about 5$\times$10$^8~$cm$^{-2}$.
Besides the well oriented NWs we see GaAs hillhocks.\cite{Breuer2011}
 During the last phase of GaAs NW growth no more Ga is supplied, and so the remaining Ga in the droplet on top of the NWs is consumed leading to a prolongation of the NW at reduced diameter. These thinner end pieces of the GaAs NWs can be recognized in Fig.~\ref{fig:SEM}~(a).
 Figure~\ref{fig:SEM}~(b) shows a  multi beam TEM micrograph of a GaAs NW. Planar defects can be recognized near the Si/GaAs IF and near the area of diameter reduction of the NW. The other regions of the NW are free of defects.

% fig.2
\begin{figure}[!t]
\includegraphics[width=14.0cm]{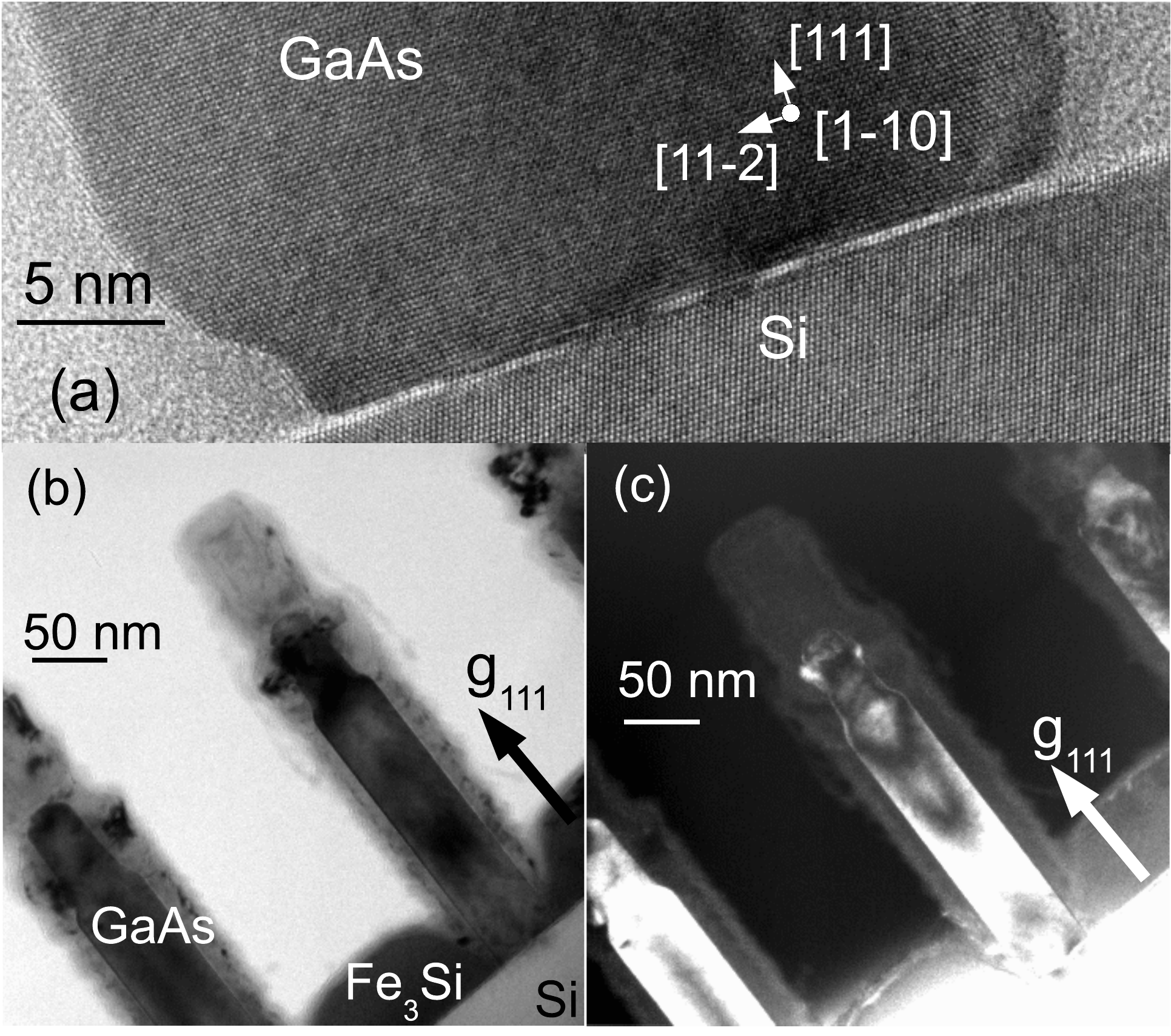}
\caption{(a) HRTEM micrograph illustrating the epitaxy of a GaAs NW on Si(111), (Sample~0). (b) TEM DF image of a GaAs-Fe$_{3}$Si core-shell NW, near the GaAs 111 reflection (Sample~1, $T_{S}$~= ~100$^\circ$C). (c) Corresponding TEM BF image of the same NW structure.}
\label{fig:BF}
\end{figure}

Figure~\ref{fig:BF}~(a) demonstrates a HRTEM micrograph of the Si/GaAs IF illustrating the epitaxial alignment of a GaAs NW on Si(111), (sample~0). No amorphous material (i.e. SiO$_2$) is observed at the GaAs/Si IF. Probably a remaining thin SiO$_2$ film was etched away  by the Ga droplet \cite{takemoto2006}.  Here, the IF is a perfect twin boundary, however NWs without twinning at the IF are observed as well.

% fig.3
\begin{figure}[!t]
\includegraphics[width=14.0cm]{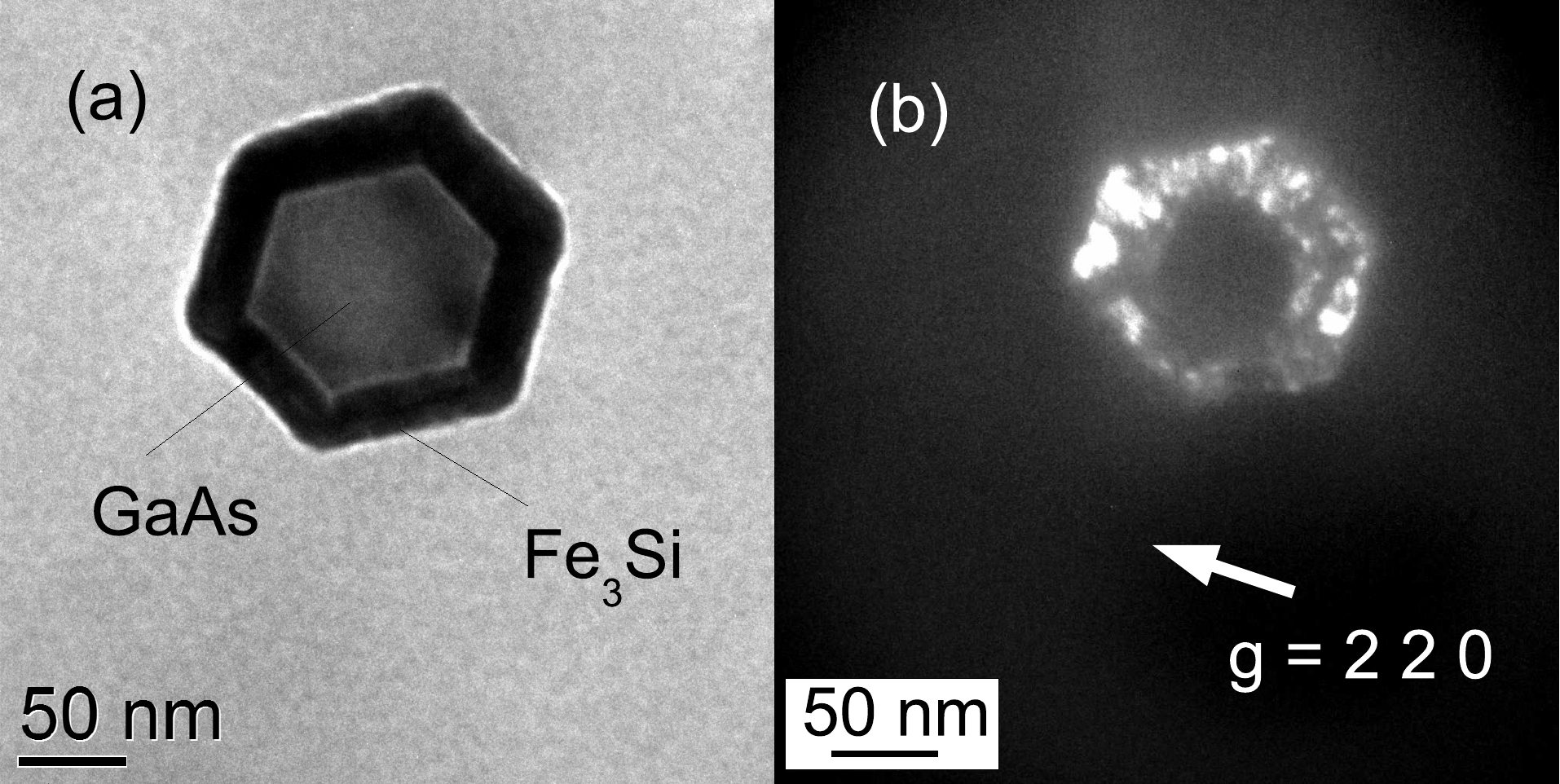}
\caption{(a) TEM BF image of a cross-section perpendicular to the axis of a GaAs-Fe$_{3}$Si core-shell NW (Sample~1, $T_{S}$~= ~100$^\circ$C).  (b) TEM DF image of a similar GaAs-Fe$_{3}$Si core-shell NW in the vicinity of (a).}
\label{fig:cross}
\end{figure}

Figure~\ref{fig:BF}~(b) displays the dark-field (DF) image of a GaAs-Fe$_{3}$Si core-shell NW taken under g = 111 (sample~1). The GaAs core is dark and the Fe$_{3}$Si shell region yields an inhomogeneous distribution of intensity, due to its textured structure. It fulfills the diffraction condition in correspondence to the different orientations of the individual crystallites. The shell is (18$\pm$3)~nm thick at the sidewall, where $\pm$3~nm characterizes the thickness inhomogeneity and not the error of the measurement. This thickness does not correspond to the equivalent amount of material deposited. It seems, that a higher amount of the Fe$_{3}$Si is integrated into more than 60~nm thick parasitic film. The material thickness on top of the NWs is even higher (about 86~nm). The value of 86~nm corresponds to the nominal film thickness of 69~nm expected for a planar structure. The diffusion of Fe$_{3}$Si along the sidewalls can be neglected, in first approximation. Then, the reduced film thickness along the sidewalls of the NWs can be explained by the small angle between the direction of the material flux and the NWs in the MBE system.
Figure~\ref{fig:BF}~(c) shows the corresponding BF image of the core-shell NW shown in (b). Here, the Fe$_{3}$Si shell does not fulfil the diffraction condition, however the single-crystalline GaAs core diffracts strongly, resulting in a high intensity of the GaAs 111 reflection. Figure~\ref{fig:BF}~(b) and (c) show clearly the reduction of the diameters of the cores after closing the Ga-shutter. Two different diameters of the GaAs cores [e.g. 50~nm and about 33~nm in Figs.~\ref{fig:BF}~(b,c)] can be recognized, evidencing two stages of NW growth.

Figure~\ref{fig:cross}(a) shows a TEM BF image of a plan-view along the axis of a GaAs-Fe$_{3}$Si core-shell NW (sample~1).
The single-crystalline GaAs core exhibits an almost homogenous contrast. The shell is (26$\pm$7)~nm thick. The thickest regions are found at the corners. We have choosen typical NWs for the determination of the thickness of the Fe$_{3}$Si shell, the variation within one NW is large compared to the differences between them.  Figure~\ref{fig:cross}(b) shows a TEM DF image of a similar GaAs-Fe$_{3}$Si core-shell NW in the vicinity of (a). Here only some parts of the Fe$_{3}$Si shell are fulfilling the diffraction condition, i.e. the shell is poly-crystalline. However, the ferromagnetic Fe$_3$Si shells cover the GaAs core NWs completely and continuously without any holes. From that point of view almost perfect ferromagnet/semiconductor structures are fabricated. The hexagonal shape of the GaAs core is reproduced by the Fe$_{3}$Si shell despite of the textured polycrystal structure of the shell.

% fig.4
\begin{figure}[!t]
\includegraphics[width=14.0cm]{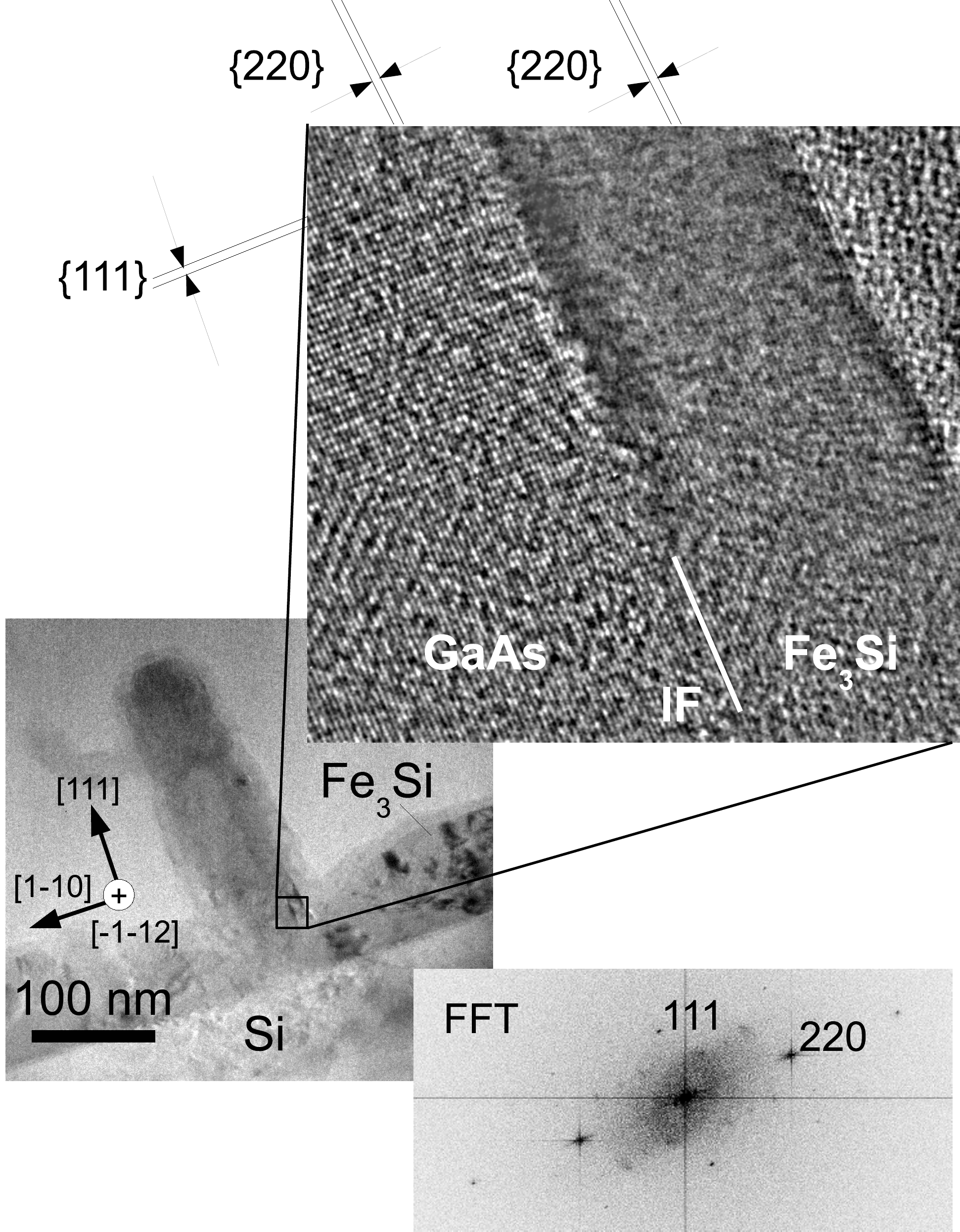}
\caption{Multi-beam TEM image of a  GaAs NW (overview and HRTEM micrograph) together with the Si substrate and the Fe$_{3}$Si shell and the parasitic Fe$_{3}$Si film of sample~1. The higher magnification is a Fourier filtered image of the marked region. The dark contrast near the IF is due to a slight tilt of the Fe$_{3}$Si lattice. A fast Fourier transform (FFT) of the HRTEM micrograph is added below (negative).
}
\label{fig:HRover}
\end{figure}

Figure~\ref{fig:HRover} reveals a multi-beam TEM image of a NW (sample~1) along the GaAs 110 zone axis together with a magnified part of the boundary in high-resolution mode. The GaAs core shows the {111} planes resolved as well as the perpendicular {220} planes, whereas only the 220 lattice planes of the Fe$_3$Si shell are depicted. The present shell crystallite seems to be slightly rotated around the [110] direction compared to the core. This finding is confirmed by the corresponding fast Fourier transform of the HRTEM micrograph shown below. The peaks of GaAs and Fe$_{3}$Si coincide due to the small lattice parameter difference of both materials.

%%These structural findings were confirmed by selected area diffraction. Sometimes the NWs contained planar defects, probably stacking faults leading to a change of their structure from zincblende to wurtzite and back.\cite{Breuer2010}

% fig.5
\begin{figure}[!t]
\includegraphics[width=14.0cm]{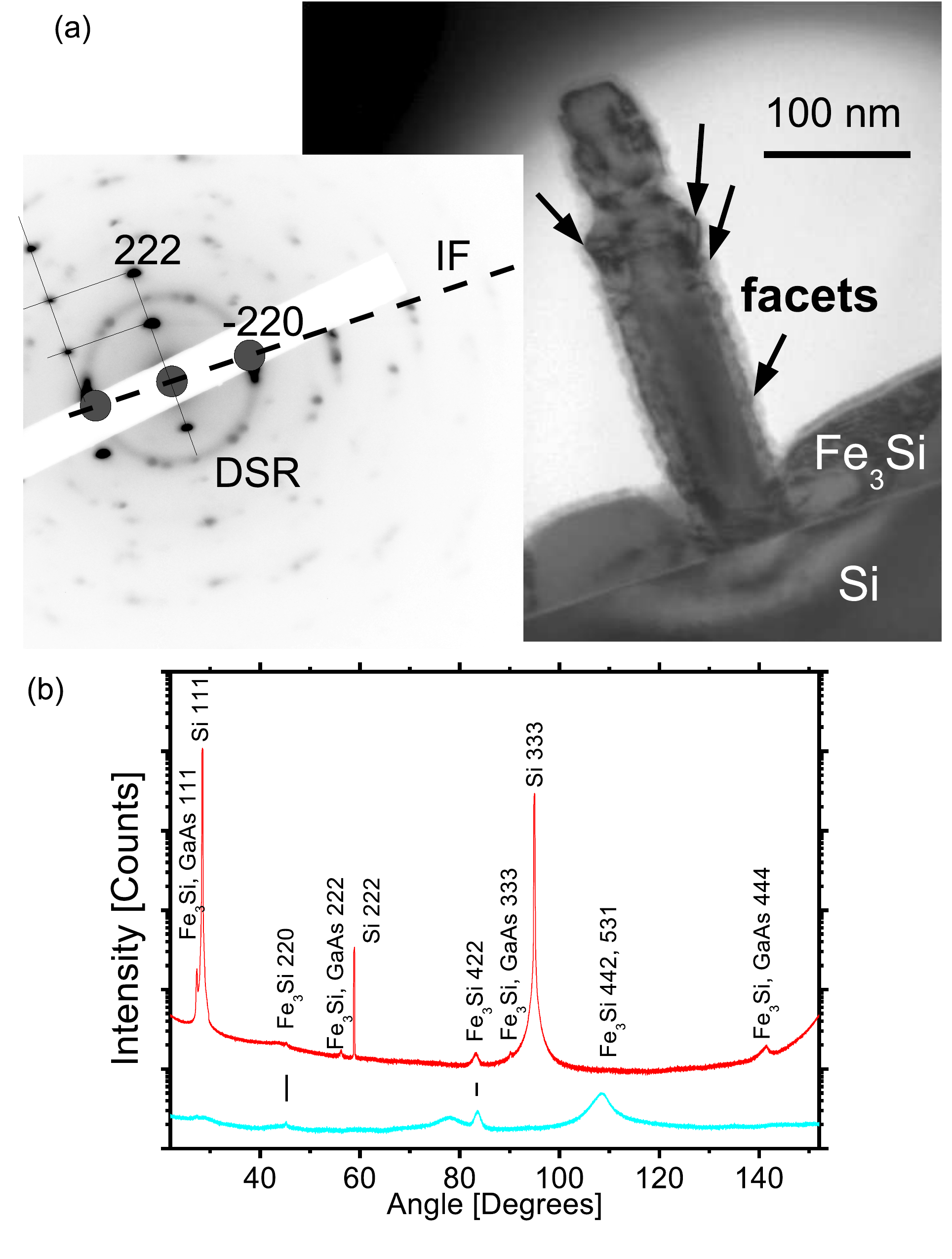}
\caption{(a) Multi-beam TEM image of a GaAs-Fe$_{3}$Si core-shell NW and corresponding electron diffraction pattern of sample~1. The growth of Fe$_{3}$Si facets is marked by arrows in the micrograph. The dashed line illustrates the orientation of the Si/GaAs interface (IF). (b) XRD pattern of the same sample. 2$\theta/\omega$-scan (upper curve) and 2$\theta$-scan (lower curve) with $\omega$=20$^\circ$ are compared.
}
\label{fig:DP1}
\end{figure}

Figure~\ref{fig:DP1}~(a) shows a multi-beam TEM image of a NW (sample~1) and the corresponding electron diffraction pattern of the NW. Surface facets of the Fe$_{3}$Si shell are marked by arrows. In the diffraction pattern the grid of spots of the GaAs NW is superimposed by ring-shaped segments coming from the Fe$_{3}$Si shell. These inhomogenous Debye-Scherrer-Rings (DSRs) are due to the textured structure of the  Fe$_{3}$Si. The dashed line illustrates the orientation of the Si/GaAs interface (IF). Figure~\ref{fig:DP1}~(b) demonstrates the XRD patterns of the same sample. The 2$\theta/\omega$-scan (upper curve) and a 2$\theta$-scan with $\omega$=20$^\circ$ (lower curve) are compared, and the 220 maximum as well as the 422 maximum of Fe$_{3}$Si appear in both scans, i.e. some parts of the Fe$_{3}$Si are oriented more or less randomly.

% fig.6
\begin{figure}[!t]
\includegraphics[width=14.0cm]{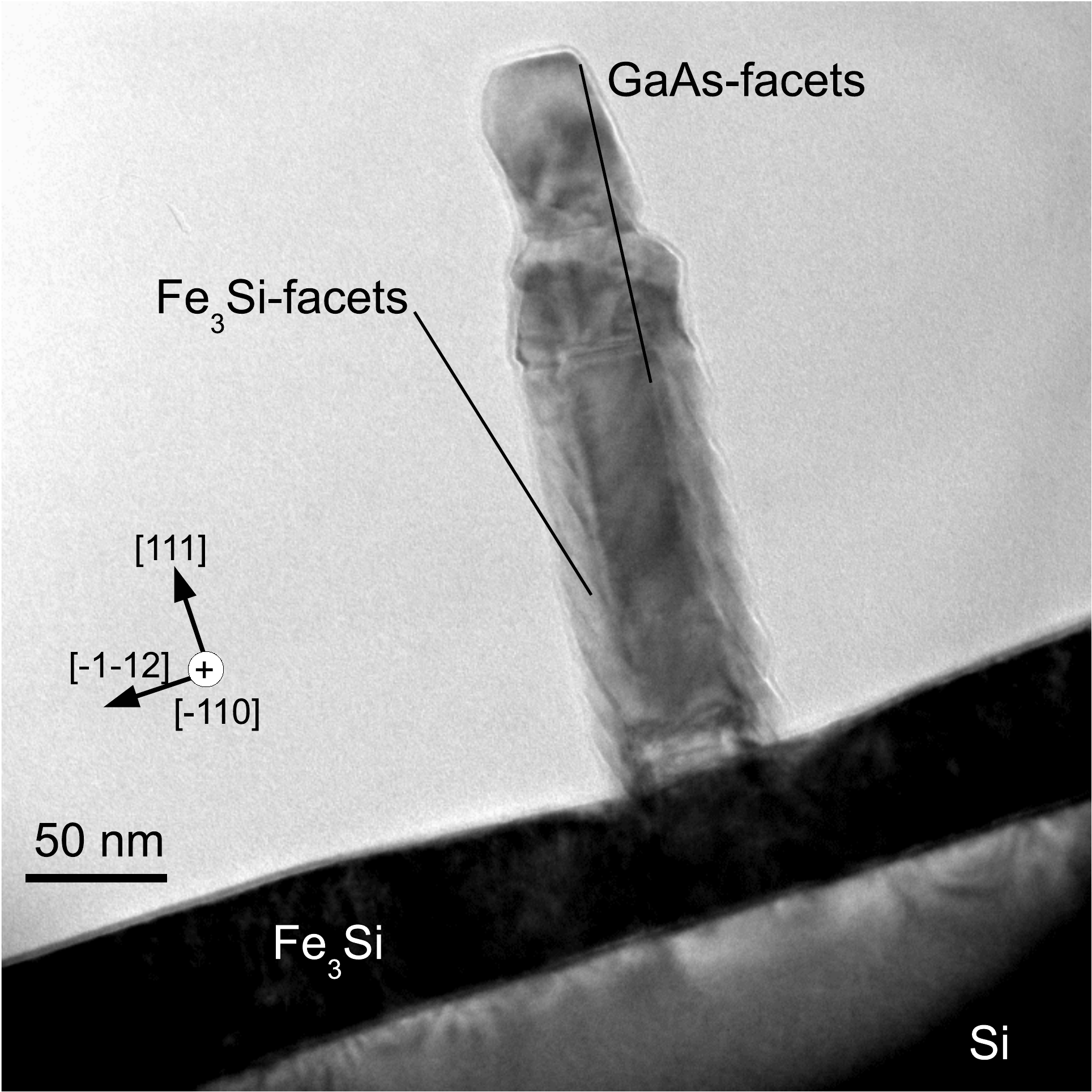}
\caption{Multi-beam TEM image of a GaAs-Fe$_{3}$Si core-shell NW (Sample~2, $T_{S}$~= ~200$^\circ$C). The growth of Fe$_{3}$Si facets and the GaAs facet are marked by lines in the micrograph. In the present micrograph the NW is much thinner than the overall sample thickness. Thus the modification of the parasitic Fe$_{3}$Si film by the NW is not obvious here, while a part of the film is covering the bottom part of the NW.
}
\label{fig:DP}
\end{figure}

Figure~\ref{fig:DP} shows a multi-beam TEM image of the NW sample~2, where the shell is grown at $T_{S}$~= ~200$^\circ$C. The growth of Fe$_{3}$Si facets is marked by lines in the micrograph. The angles between the facets of the GaAs NWs and of the Fe$_{3}$Si shells are smaller than those of the other samples. In this way the Fe$_{3}$Si surface seems to be the smoothest.

% fig.7
\begin{figure}[!t]
\includegraphics[width=9.0cm]{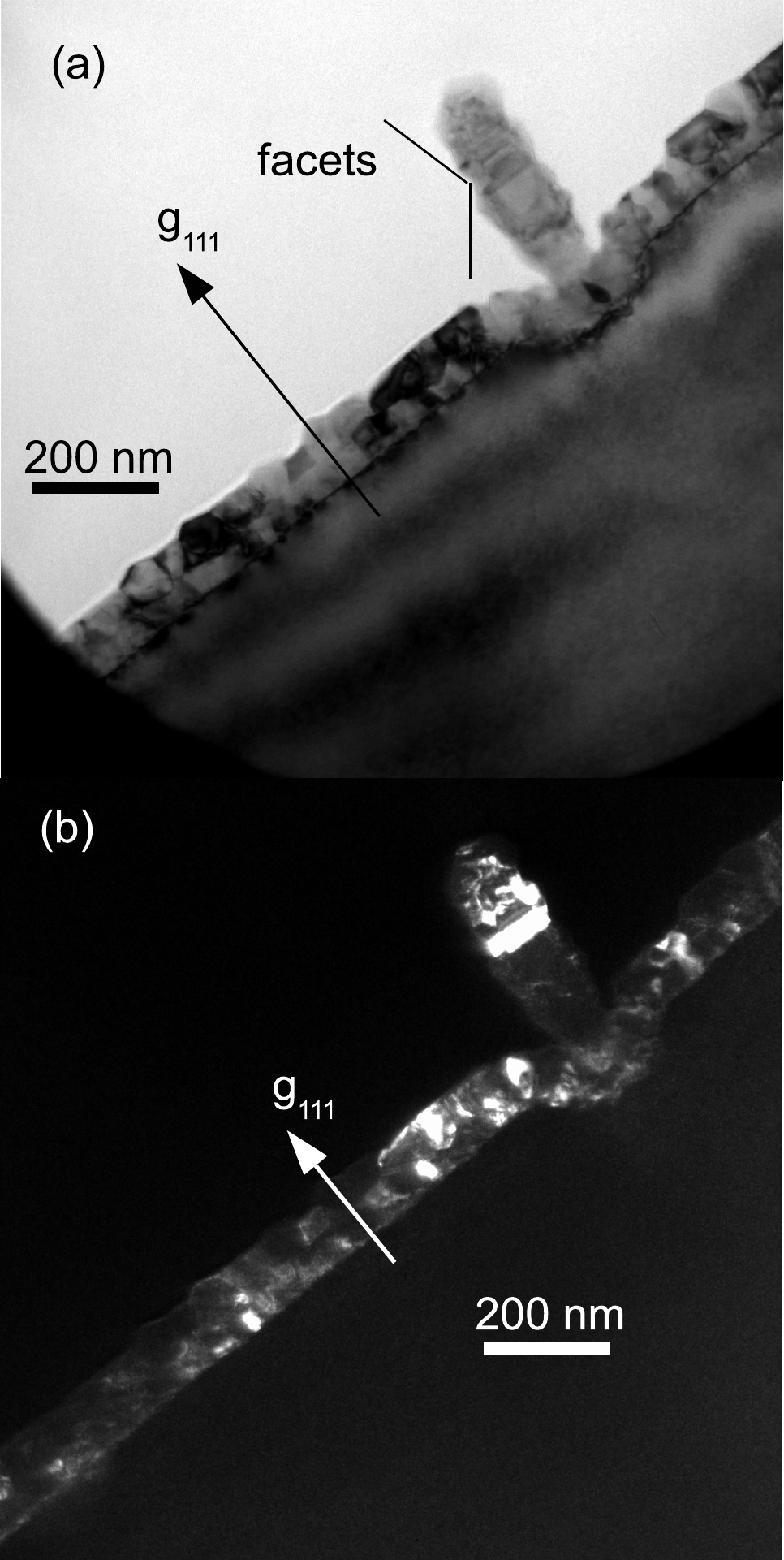}

\caption{(a) BF TEM image of a  GaAs-Fe$_{3}$Si core-shell NW (Sample~3, $T_{S}$~= ~350$^\circ$C). The growth of facets is marked by lines. (b) DF TEM image of a GaAs-Fe$_{3}$Si core shell NW together with the Fe$_{3}$Si film.
}
\label{fig:DFht}
\end{figure}

% fig.8
\begin{figure}[!t]
\includegraphics[width=12.0cm]{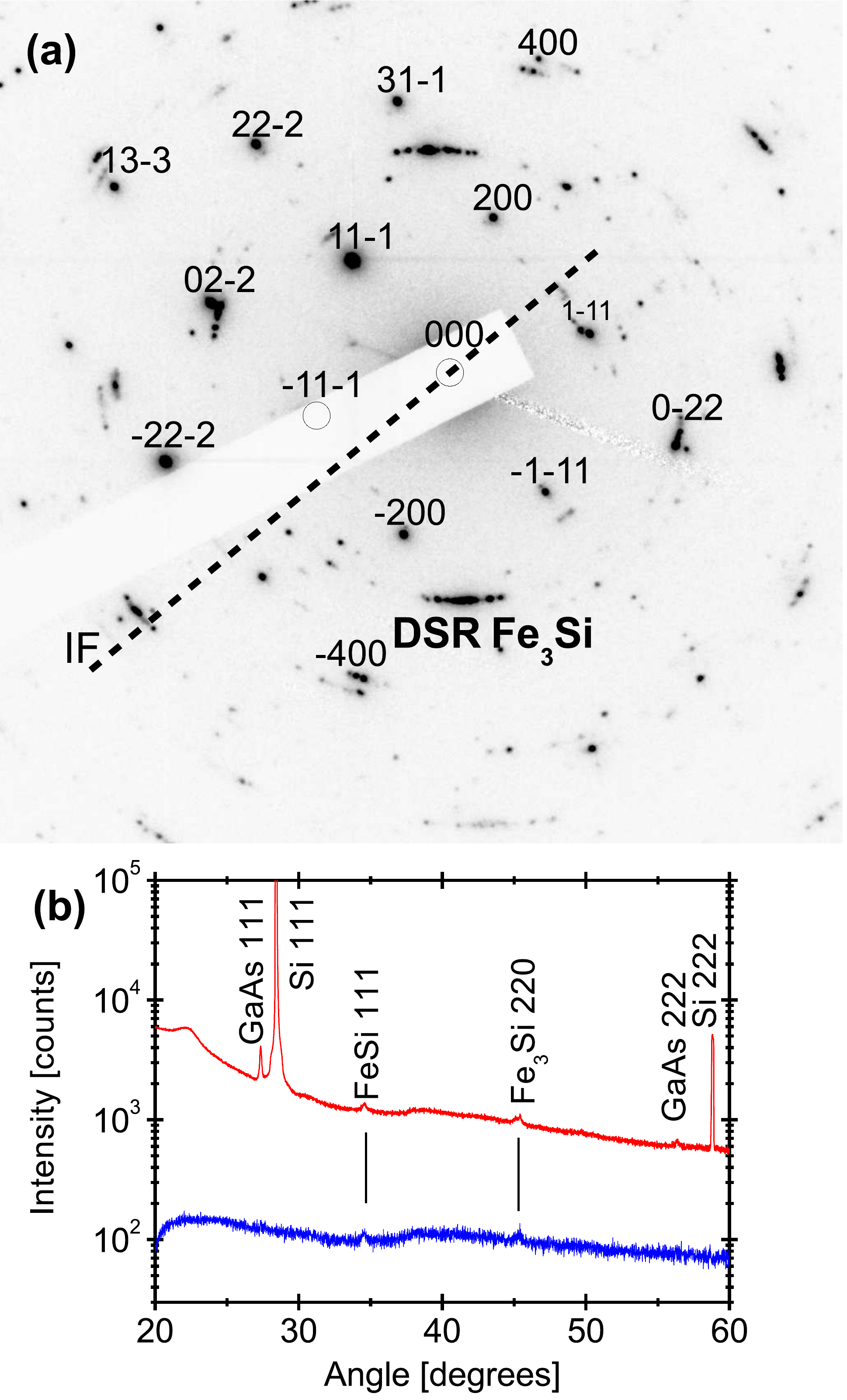}
\caption{(a) Selected area electron diffraction pattern of the region of sample~3 corresponding to Fig.~\ref{fig:DFht}. Some of the diffraction spots form pieces of DSRs. The dashed line illustrates the orientation of the IF. (b) XRD-curves of sample~3. The upper curve is an $2\Theta/\omega$-scan whereas the lower curve is $2\Theta$-scan with a constant angle of incidence of 20$^\circ$.
}
\label{fig:DFhtXRD}
\end{figure}

Figure~\ref{fig:DFht}~(a) shows a BF TEM image of a typical GaAs-Fe$_{3}$Si core-shell NW, where the shell was grown at an even higher substrate temperature of $T_{S}$~= ~350$^\circ$C (sample~3). The inhomogeneity of the shell image is present in a similar manner as for sample~1. However, facets are more pronounced. The shell thickness in this NW is (33$\pm$15)~nm at the sidewalls, and (90$\pm$5)~nm on top of the NW (not shown here). The parasitic Fe$_{3}$Si film is (105$\pm$15)~nm thick. In addition, a severe reaction of the Fe$_{3}$Si with the Si substrate has occurred near the NW resulting in a degradation of the interface and the formation of a 60~nm deep crater around the NW. The depth of the craters on sample~3 is ranging between 50 and 100~nm. Melt back etching of the Si surface near Ga droplets at high temperatures was observed earlier. \cite{Pinzone1987,Sakai1988,takemoto2006,Zhu2013} In our view a similar kind of reaction could be the reason for the formation of the craters. Precipitates were formed on the substrate side of the IF as well.
Figure~\ref{fig:DFht}~(b) demonstrates the TEM DF  image of the NW together with the Fe$_{3}$Si film (sample~3).  Grainy bright contrasts of the film and the  NW shell depict their poly-crystalline structure. {111} oriented crystallites appear bright in the micrograph. Figure~\ref{fig:DFhtXRD}~(a) demonstrates the corresponding SAD  pattern. Sections of DSRs are evidencing the textured structure of the Fe$_{3}$Si. The ring segments of the SAD pattern are corresponding to the net-plane distances of Fe$_{3}$Si 220, 004, and 133. The DSRs consist of several spots, indicating that the orientation distribution is not homogeneous (texture).
Figure~\ref{fig:DFhtXRD}~(b) shows XRD-curves of sample~3. The upper curve is a $2\Theta/\omega$-scan whereas the lower curve is $2\Theta$-scan with a constant angle of incidence of 20$^\circ$. The Fe$_{3}$Si 220 reflection and the FeSi 111 reflection occur in both scans, i.e. the Fe$_{3}$Si and the FeSi  are poly-crystalline. Hence, the FeSi alloy is not a single phase any more. The Si and GaAs reflections occur only in the upper scan, i.e. Si and GaAs are single crystalline. As the reflections of GaAs and Fe$_{3}$Si practically coincide, diffraction of some [111] oriented Fe$_{3}$Si crystallites may be contained in the GaAs reflections as well.

We observe a clear tendency of stronger Fe$_{3}$Si facetting with increasing T$_S$ [cf. Figs.~\ref{fig:DP}~(a) and \ref{fig:DFht}~(a), e.g.]. Usually these facets are observed to be inclined to the NW axis. The formation of facets reduces the overall surface energy and evidences non negligible material transport over distances small compared to the NW lengths. Unfortunately, the orientation of those facets does not coincide with those of the original GaAs NWs. That is why the shell thickness inhomogeneity is increasing with T$_S$ due to the facetted growth.
Higher substrate temperatures during growth of the Fe$_{3}$Si shell lead to enhanced surface diffusion. Under the condition of poor wetting the enhanced diffusion leads to formation of Fe$_{3}$Si islands having an equilibrium shape, i.e. their own facets. At even higher substrate temperatures we found chemical reactions at the Fe$_{3}$Si/GaAs and the GaAs/Si interfaces causing the extended pits around the core-shell NWs together with the partial destruction of the GaAs cores.

Layer-by-layer growth at medium substrate temperatures could in principle solve the problem of facetting. However, even planar Fe$_3$Si grows on GaAs in the Vollmer-Weber (VW) growth mode \cite{kag09}. Poor wetting during the growth of Fe$_3$Si on the GaAs surface leads initially to isolated islands, although both lattices match perfectly.  We speculate that textured  Fe$_3$Si shells found in the present work are a result of the VW growth mode. In general, one way to improve the homogeneity and the other structural properties of the Fe$_{3}$Si-shells could be the use of surfactants.~\cite{Copel1989,LeGoues1990,hoegen1991,Tromp1992,vegt1992}
For the growth on GaAs the following surfactant materials are under discussion: Sb~\cite{Kageyama2004}, Bi~\cite{Tixier2003}, Te~\cite{Grandjean1992}, Pb~\cite{Massies1993}. The surfactant Sb  improved the growth of  other silicides as well.~\cite{Teichert2002}

\section{Conclusions}

The GaAs core NWs grow epitaxially on the bare Si(111) surface inside holes of the SiO$_2$ film via the VLS growth mechanism. A GaAs/Si IF is formed without any amorphous phase in between. Two stages of the growth of the GaAs cores can be distinguished: regular growth and residual growth after ending of the Ga supply. The surfaces of the original GaAs NWs are completely covered by magnetic Fe$_3$Si exhibiting an enhanced surface roughness compared to the bare GaAs NWs, partly due to formation of facets. Nevertheless, continuous magnetic shells are established.
Bright- and dark-field micrographs reveal homogeneous contrast in the GaAs cores and pronounced inhomogeneity in the Fe$_3$Si shells and the parasitic Fe$_3$Si film.
The substrate temperature during Fe$_{3}$Si deposition has a strong influence on the structural quality of the NW structures, especially the Fe$_{3}$Si shells.

Growth of the shells at high substrate temperatures above $T_{S}$~= ~200$^\circ$C leads to severe reactions between the shell and the core as well as the parasitic layer and the substrate which are detrimental for device applications.

Increased facetting is observed with growing substrate temperature T$_S$.
Poly-crystalline  Fe$_3$Si shells found in the present work are probably the result of the VW growth mode of Fe$_3$Si on the surfaces of the GaAs NWs.
Nevertheless, such a grainy structure may even be sufficient for spintronic devices, because they are forming a dense film of magnetic material around the semiconducting cores.

\section{Acknowledgement}
The authors thank Claudia Herrmann, for her support during the
MBE growth, Doreen Steffen for sample preparation, Astrid Pfeiffer
for help in the laboratory,  Anne-Kathrin Bluhm for the SEM micrographs, Esperanza Luna and Uwe Jahn
for valuable support and helpful discussion.

\section{References}

%\bibliography{Zitate}

\begin{thebibliography}{41}%
\makeatletter
\providecommand \@ifxundefined [1]{%
 \@ifx{#1\undefined}
}%
\providecommand \@ifnum [1]{%
 \ifnum #1\expandafter \@firstoftwo
 \else \expandafter \@secondoftwo
 \fi
}%
\providecommand \@ifx [1]{%
 \ifx #1\expandafter \@firstoftwo
 \else \expandafter \@secondoftwo
 \fi
}%
\providecommand \natexlab [1]{#1}%
\providecommand \enquote  [1]{``#1''}%
\providecommand \bibnamefont  [1]{#1}%
\providecommand \bibfnamefont [1]{#1}%
\providecommand \citenamefont [1]{#1}%
\providecommand \href@noop [0]{\@secondoftwo}%
\providecommand \href [0]{\begingroup \@sanitize@url \@href}%
\providecommand \@href[1]{\@@startlink{#1}\@@href}%
\providecommand \@@href[1]{\endgroup#1\@@endlink}%
\providecommand \@sanitize@url [0]{\catcode `\\12\catcode `\$12\catcode
  `\&12\catcode `\#12\catcode `\^12\catcode `\_12\catcode `\%12\relax}%
\providecommand \@@startlink[1]{}%
\providecommand \@@endlink[0]{}%
\providecommand \url  [0]{\begingroup\@sanitize@url \@url }%
\providecommand \@url [1]{\endgroup\@href {#1}{\urlprefix }}%
\providecommand \urlprefix  [0]{URL }%
\providecommand \Eprint [0]{\href }%
\providecommand \doibase [0]{http://dx.doi.org/}%
\providecommand \selectlanguage [0]{\@gobble}%
\providecommand \bibinfo  [0]{\@secondoftwo}%
\providecommand \bibfield  [0]{\@secondoftwo}%
\providecommand \translation [1]{[#1]}%
\providecommand \BibitemOpen [0]{}%
\providecommand \bibitemStop [0]{}%
\providecommand \bibitemNoStop [0]{.\EOS\space}%
\providecommand \EOS [0]{\spacefactor3000\relax}%
\providecommand \BibitemShut  [1]{\csname bibitem#1\endcsname}%
\let\auto@bib@innerbib\@empty
%</preamble>
\bibitem [{\citenamefont {Lu}\ and\ \citenamefont {Lieber}(2006)}]{lu2006}%
  \BibitemOpen
  \bibfield  {author} {\bibinfo {author} {\bibfnamefont {W.}~\bibnamefont
  {Lu}}\ and\ \bibinfo {author} {\bibfnamefont {M.}~\bibnamefont {Lieber}},\
  }\href@noop {} {\bibfield  {journal} {\bibinfo  {journal} {J. Phys. D: Appl.
  Phys.}\ }\textbf {\bibinfo {volume} {39}},\ \bibinfo {pages} {387} (\bibinfo
  {year} {2006})}\BibitemShut {NoStop}%
\bibitem [{\citenamefont {Lieber}\ and\ \citenamefont
  {Wang}(2007)}]{Lieber2007}%
  \BibitemOpen
  \bibfield  {author} {\bibinfo {author} {\bibfnamefont {C.~M.}\ \bibnamefont
  {Lieber}}\ and\ \bibinfo {author} {\bibfnamefont {Z.~L.}\ \bibnamefont
  {Wang}},\ }\href@noop {} {\bibfield  {journal} {\bibinfo  {journal} {MRS
  Bull.}\ }\textbf {\bibinfo {volume} {32}},\ \bibinfo {pages} {99} (\bibinfo
  {year} {2007})}\BibitemShut {NoStop}%
\bibitem [{\citenamefont {Couto}\ \emph {et~al.}(2012)\citenamefont {Couto},
  \citenamefont {Sercombe}, \citenamefont {Puebla}, \citenamefont {Otubo},
  \citenamefont {Luxmoore}, \citenamefont {Sich}, \citenamefont {Elliot},
  \citenamefont {Chekhovich}, \citenamefont {Wilson}, \citenamefont {Skolnick},
  \citenamefont {Liu},\ and\ \citenamefont {Tartakovskii}}]{couto12}%
  \BibitemOpen
  \bibfield  {author} {\bibinfo {author} {\bibfnamefont {O.~D.~D.}\
  \bibnamefont {Couto}}, \bibinfo {author} {\bibfnamefont {D.}~\bibnamefont
  {Sercombe}}, \bibinfo {author} {\bibfnamefont {J.}~\bibnamefont {Puebla}},
  \bibinfo {author} {\bibfnamefont {L.}~\bibnamefont {Otubo}}, \bibinfo
  {author} {\bibfnamefont {I.~J.}\ \bibnamefont {Luxmoore}}, \bibinfo {author}
  {\bibfnamefont {M.}~\bibnamefont {Sich}}, \bibinfo {author} {\bibfnamefont
  {T.~J.}\ \bibnamefont {Elliot}}, \bibinfo {author} {\bibfnamefont {E.~A.}\
  \bibnamefont {Chekhovich}}, \bibinfo {author} {\bibfnamefont {L.~R.}\
  \bibnamefont {Wilson}}, \bibinfo {author} {\bibfnamefont {M.~S.}\
  \bibnamefont {Skolnick}}, \bibinfo {author} {\bibfnamefont {H.~Y.}\
  \bibnamefont {Liu}}, \ and\ \bibinfo {author} {\bibfnamefont {A.~I.}\
  \bibnamefont {Tartakovskii}},\ }\href@noop {} {\bibfield  {journal} {\bibinfo
   {journal} {Nano Letters}\ }\textbf {\bibinfo {volume} {12}},\ \bibinfo
  {pages} {5269} (\bibinfo {year} {2012})}\BibitemShut {NoStop}%
\bibitem [{\citenamefont {Breuer}\ \emph {et~al.}(2011)\citenamefont {Breuer},
  \citenamefont {Pfueller}, \citenamefont {Flissikowski}, \citenamefont
  {Brandt}, \citenamefont {Grahn}, \citenamefont {Geelhaar},\ and\
  \citenamefont {Riechert}}]{Breuer2011}%
  \BibitemOpen
  \bibfield  {author} {\bibinfo {author} {\bibfnamefont {S.}~\bibnamefont
  {Breuer}}, \bibinfo {author} {\bibfnamefont {C.}~\bibnamefont {Pfueller}},
  \bibinfo {author} {\bibfnamefont {T.}~\bibnamefont {Flissikowski}}, \bibinfo
  {author} {\bibfnamefont {O.}~\bibnamefont {Brandt}}, \bibinfo {author}
  {\bibfnamefont {H.~T.}\ \bibnamefont {Grahn}}, \bibinfo {author}
  {\bibfnamefont {L.}~\bibnamefont {Geelhaar}}, \ and\ \bibinfo {author}
  {\bibfnamefont {H.}~\bibnamefont {Riechert}},\ }\href@noop {} {\bibfield
  {journal} {\bibinfo  {journal} {Nano Lett.}\ }\textbf {\bibinfo {volume}
  {11}},\ \bibinfo {pages} {1276} (\bibinfo {year} {2011})}\BibitemShut
  {NoStop}%
\bibitem [{\citenamefont {Hilse}\ \emph {et~al.}(2009)\citenamefont {Hilse},
  \citenamefont {Takagaki}, \citenamefont {Herfort}, \citenamefont
  {Ramsteiner}, \citenamefont {Herrmann}, \citenamefont {Breuer}, \citenamefont
  {Geelhaar},\ and\ \citenamefont {Riechert}}]{Hilse2009}%
  \BibitemOpen
  \bibfield  {author} {\bibinfo {author} {\bibfnamefont {M.}~\bibnamefont
  {Hilse}}, \bibinfo {author} {\bibfnamefont {Y.}~\bibnamefont {Takagaki}},
  \bibinfo {author} {\bibfnamefont {J.}~\bibnamefont {Herfort}}, \bibinfo
  {author} {\bibfnamefont {M.}~\bibnamefont {Ramsteiner}}, \bibinfo {author}
  {\bibfnamefont {C.}~\bibnamefont {Herrmann}}, \bibinfo {author}
  {\bibfnamefont {S.}~\bibnamefont {Breuer}}, \bibinfo {author} {\bibfnamefont
  {L.}~\bibnamefont {Geelhaar}}, \ and\ \bibinfo {author} {\bibfnamefont
  {H.}~\bibnamefont {Riechert}},\ }\href@noop {} {\bibfield  {journal}
  {\bibinfo  {journal} {Appl. Phys. Lett.}\ }\textbf {\bibinfo {volume} {95}},\
  \bibinfo {pages} {133126} (\bibinfo {year} {2009})}\BibitemShut {NoStop}%
\bibitem [{\citenamefont {Rudolph}\ \emph {et~al.}(2009)\citenamefont
  {Rudolph}, \citenamefont {Soda}, \citenamefont {Kiessling}, \citenamefont
  {Wojtowicz}, \citenamefont {Schuh}, \citenamefont {Wegscheider},
  \citenamefont {Zweck}, \citenamefont {Back},\ and\ \citenamefont
  {Reiger}}]{Rudolph2009}%
  \BibitemOpen
  \bibfield  {author} {\bibinfo {author} {\bibfnamefont {A.}~\bibnamefont
  {Rudolph}}, \bibinfo {author} {\bibfnamefont {M.}~\bibnamefont {Soda}},
  \bibinfo {author} {\bibfnamefont {M.}~\bibnamefont {Kiessling}}, \bibinfo
  {author} {\bibfnamefont {T.}~\bibnamefont {Wojtowicz}}, \bibinfo {author}
  {\bibfnamefont {D.}~\bibnamefont {Schuh}}, \bibinfo {author} {\bibfnamefont
  {W.}~\bibnamefont {Wegscheider}}, \bibinfo {author} {\bibfnamefont
  {J.}~\bibnamefont {Zweck}}, \bibinfo {author} {\bibfnamefont
  {C.}~\bibnamefont {Back}}, \ and\ \bibinfo {author} {\bibfnamefont
  {E.}~\bibnamefont {Reiger}},\ }\href@noop {} {\bibfield  {journal} {\bibinfo
  {journal} {Nano Lett.}\ }\textbf {\bibinfo {volume} {9}},\ \bibinfo {pages}
  {3860} (\bibinfo {year} {2009})}\BibitemShut {NoStop}%
\bibitem [{\citenamefont {Rueffer}\ \emph {et~al.}(2012)\citenamefont
  {Rueffer}, \citenamefont {Huber},\ and\ \citenamefont
  {Berberich}}]{Rueffer2012}%
  \BibitemOpen
  \bibfield  {author} {\bibinfo {author} {\bibfnamefont {D.}~\bibnamefont
  {Rueffer}}, \bibinfo {author} {\bibfnamefont {R.}~\bibnamefont {Huber}}, \
  and\ \bibinfo {author} {\bibfnamefont {P.}~\bibnamefont {Berberich}},\
  }\href@noop {} {\bibfield  {journal} {\bibinfo  {journal} {Nanoscale}\
  }\textbf {\bibinfo {volume} {4}},\ \bibinfo {pages} {4989} (\bibinfo {year}
  {2012})}\BibitemShut {NoStop}%
\bibitem [{\citenamefont {Dellas}\ \emph {et~al.}(2010)\citenamefont {Dellas},
  \citenamefont {Liang}, \citenamefont {Cooley}, \citenamefont {Samarth},\ and\
  \citenamefont {Mohney}}]{Dellas2010}%
  \BibitemOpen
  \bibfield  {author} {\bibinfo {author} {\bibfnamefont {N.~S.}\ \bibnamefont
  {Dellas}}, \bibinfo {author} {\bibfnamefont {J.}~\bibnamefont {Liang}},
  \bibinfo {author} {\bibfnamefont {B.~J.}\ \bibnamefont {Cooley}}, \bibinfo
  {author} {\bibfnamefont {N.}~\bibnamefont {Samarth}}, \ and\ \bibinfo
  {author} {\bibfnamefont {S.~E.}\ \bibnamefont {Mohney}},\ }\href@noop {}
  {\bibfield  {journal} {\bibinfo  {journal} {Appl. Phys. Lett.}\ }\textbf
  {\bibinfo {volume} {97}},\ \bibinfo {pages} {072505} (\bibinfo {year}
  {2010})}\BibitemShut {NoStop}%
\bibitem [{\citenamefont {Tivakornsasithorn}\ \emph {et~al.}(2012)\citenamefont
  {Tivakornsasithorn}, \citenamefont {Pimpinella}, \citenamefont {Nguyen},
  \citenamefont {Liu}, \citenamefont {Dobrowolska},\ and\ \citenamefont
  {Furdyna}}]{Tivakorn2012}%
  \BibitemOpen
  \bibfield  {author} {\bibinfo {author} {\bibfnamefont {K.}~\bibnamefont
  {Tivakornsasithorn}}, \bibinfo {author} {\bibfnamefont {R.~E.}\ \bibnamefont
  {Pimpinella}}, \bibinfo {author} {\bibfnamefont {V.}~\bibnamefont {Nguyen}},
  \bibinfo {author} {\bibfnamefont {X.}~\bibnamefont {Liu}}, \bibinfo {author}
  {\bibfnamefont {M.}~\bibnamefont {Dobrowolska}}, \ and\ \bibinfo {author}
  {\bibfnamefont {J.~K.~J.}\ \bibnamefont {Furdyna}},\ }\href@noop {}
  {\bibfield  {journal} {\bibinfo  {journal} {J. Vac. Sci. Technol. B}\
  }\textbf {\bibinfo {volume} {30}},\ \bibinfo {pages} {02115} (\bibinfo {year}
  {2012})}\BibitemShut {NoStop}%
\bibitem [{\citenamefont {Yu}\ \emph {et~al.}(2013)\citenamefont {Yu},
  \citenamefont {Wang}, \citenamefont {Pan}, \citenamefont {Zhao},
  \citenamefont {Misuraca}, \citenamefont {Molnar},\ and\ \citenamefont
  {Xiong}}]{Yu2013}%
  \BibitemOpen
  \bibfield  {author} {\bibinfo {author} {\bibfnamefont {X.}~\bibnamefont
  {Yu}}, \bibinfo {author} {\bibfnamefont {H.}~\bibnamefont {Wang}}, \bibinfo
  {author} {\bibfnamefont {D.}~\bibnamefont {Pan}}, \bibinfo {author}
  {\bibfnamefont {J.}~\bibnamefont {Zhao}}, \bibinfo {author} {\bibfnamefont
  {J.}~\bibnamefont {Misuraca}}, \bibinfo {author} {\bibfnamefont
  {S.}~\bibnamefont {Molnar}}, \ and\ \bibinfo {author} {\bibfnamefont
  {P.}~\bibnamefont {Xiong}},\ }\href@noop {} {\bibfield  {journal} {\bibinfo
  {journal} {Nano Lett.}\ }\textbf {\bibinfo {volume} {13}},\ \bibinfo {pages}
  {1572} (\bibinfo {year} {2013})}\BibitemShut {NoStop}%
\bibitem [{\citenamefont {Farshchi}\ \emph {et~al.}(2011)\citenamefont
  {Farshchi}, \citenamefont {Ramsteiner}, \citenamefont {Herfort},
  \citenamefont {Tahraoui},\ and\ \citenamefont {Grahn}}]{Farshchi2011}%
  \BibitemOpen
  \bibfield  {author} {\bibinfo {author} {\bibfnamefont {R.}~\bibnamefont
  {Farshchi}}, \bibinfo {author} {\bibfnamefont {M.}~\bibnamefont
  {Ramsteiner}}, \bibinfo {author} {\bibfnamefont {J.}~\bibnamefont {Herfort}},
  \bibinfo {author} {\bibfnamefont {A.}~\bibnamefont {Tahraoui}}, \ and\
  \bibinfo {author} {\bibfnamefont {H.~T.}\ \bibnamefont {Grahn}},\ }\href@noop
  {} {\bibfield  {journal} {\bibinfo  {journal} {Appl. Phys. Lett.}\ }\textbf
  {\bibinfo {volume} {98}},\ \bibinfo {pages} {162508} (\bibinfo {year}
  {2011})}\BibitemShut {NoStop}%
\bibitem [{\citenamefont {Parkin}\ \emph {et~al.}(2008)\citenamefont {Parkin},
  \citenamefont {Hayashi},\ and\ \citenamefont {Thomas}}]{Parkin2008}%
  \BibitemOpen
  \bibfield  {author} {\bibinfo {author} {\bibfnamefont {S.~S.}\ \bibnamefont
  {Parkin}}, \bibinfo {author} {\bibfnamefont {M.}~\bibnamefont {Hayashi}}, \
  and\ \bibinfo {author} {\bibfnamefont {L.}~\bibnamefont {Thomas}},\
  }\href@noop {} {\bibfield  {journal} {\bibinfo  {journal} {Science}\ }\textbf
  {\bibinfo {volume} {320}},\ \bibinfo {pages} {190} (\bibinfo {year}
  {2008})}\BibitemShut {NoStop}%
\bibitem [{\citenamefont {Ryu}\ \emph {et~al.}(2012)\citenamefont {Ryu},
  \citenamefont {Thomas}, \citenamefont {Yang},\ and\ \citenamefont
  {Parkin}}]{Ryu2012}%
  \BibitemOpen
  \bibfield  {author} {\bibinfo {author} {\bibfnamefont {K.~S.}\ \bibnamefont
  {Ryu}}, \bibinfo {author} {\bibfnamefont {L.}~\bibnamefont {Thomas}},
  \bibinfo {author} {\bibfnamefont {S.~H.}\ \bibnamefont {Yang}}, \ and\
  \bibinfo {author} {\bibfnamefont {S.~S.}\ \bibnamefont {Parkin}},\
  }\href@noop {} {\bibfield  {journal} {\bibinfo  {journal} {Appl. Phys. Exp.}\
  }\textbf {\bibinfo {volume} {5}},\ \bibinfo {pages} {093006} (\bibinfo {year}
  {2012})}\BibitemShut {NoStop}%
\bibitem [{\citenamefont {Jenichen}\ \emph {et~al.}(2005)\citenamefont
  {Jenichen}, \citenamefont {Kaganer}, \citenamefont {Herfort}, \citenamefont
  {Satapathy}, \citenamefont {Sch\"onherr}, \citenamefont {Braun},\ and\
  \citenamefont {Ploog}}]{Jenichen05}%
  \BibitemOpen
  \bibfield  {author} {\bibinfo {author} {\bibfnamefont {B.}~\bibnamefont
  {Jenichen}}, \bibinfo {author} {\bibfnamefont {V.~M.}\ \bibnamefont
  {Kaganer}}, \bibinfo {author} {\bibfnamefont {J.}~\bibnamefont {Herfort}},
  \bibinfo {author} {\bibfnamefont {D.~K.}\ \bibnamefont {Satapathy}}, \bibinfo
  {author} {\bibfnamefont {H.~P.}\ \bibnamefont {Sch\"onherr}}, \bibinfo
  {author} {\bibfnamefont {W.}~\bibnamefont {Braun}}, \ and\ \bibinfo {author}
  {\bibfnamefont {K.~H.}\ \bibnamefont {Ploog}},\ }\href@noop {} {\bibfield
  {journal} {\bibinfo  {journal} {Phys. Rev. B}\ }\textbf {\bibinfo {volume}
  {72}},\ \bibinfo {pages} {075329} (\bibinfo {year} {2005})}\BibitemShut
  {NoStop}%
\bibitem [{\citenamefont {Herfort}\ \emph {et~al.}(2012)\citenamefont
  {Herfort}, \citenamefont {Jenichen}, \citenamefont {Kaganer}, \citenamefont
  {Trampert}, \citenamefont {Schoenherr},\ and\ \citenamefont
  {Ploog}}]{Herfort2006}%
  \BibitemOpen
  \bibfield  {author} {\bibinfo {author} {\bibfnamefont {J.}~\bibnamefont
  {Herfort}}, \bibinfo {author} {\bibfnamefont {B.}~\bibnamefont {Jenichen}},
  \bibinfo {author} {\bibfnamefont {V.}~\bibnamefont {Kaganer}}, \bibinfo
  {author} {\bibfnamefont {A.}~\bibnamefont {Trampert}}, \bibinfo {author}
  {\bibfnamefont {H.~P.}\ \bibnamefont {Schoenherr}}, \ and\ \bibinfo {author}
  {\bibfnamefont {K.}~\bibnamefont {Ploog}},\ }\href@noop {} {\bibfield
  {journal} {\bibinfo  {journal} {Physica E}\ }\textbf {\bibinfo {volume}
  {32}},\ \bibinfo {pages} {371} (\bibinfo {year} {2012})}\BibitemShut
  {NoStop}%
\bibitem [{\citenamefont {Herfort}\ \emph {et~al.}(2006)\citenamefont
  {Herfort}, \citenamefont {Trampert},\ and\ \citenamefont
  {Ploog}}]{Herfort2006b}%
  \BibitemOpen
  \bibfield  {author} {\bibinfo {author} {\bibfnamefont {J.}~\bibnamefont
  {Herfort}}, \bibinfo {author} {\bibfnamefont {A.}~\bibnamefont {Trampert}}, \
  and\ \bibinfo {author} {\bibfnamefont {K.}~\bibnamefont {Ploog}},\
  }\href@noop {} {\bibfield  {journal} {\bibinfo  {journal} {Int. J. Mater.
  Res.}\ }\textbf {\bibinfo {volume} {97}},\ \bibinfo {pages} {1026} (\bibinfo
  {year} {2006})}\BibitemShut {NoStop}%
\bibitem [{\citenamefont {Hansen}(1958)}]{Hansen1958}%
  \BibitemOpen
  \bibfield  {author} {\bibinfo {author} {\bibfnamefont {M.}~\bibnamefont
  {Hansen}},\ }\href@noop {} {\emph {\bibinfo {title} {Constitution of Binary
  Alloys}}}\ (\bibinfo  {publisher} {McGraw-Hill: New York},\ \bibinfo
  {address} {San Diego, CA},\ \bibinfo {year} {1958})\BibitemShut {NoStop}%
\bibitem [{\citenamefont {Elliot}(1965)}]{Elliot1965}%
  \BibitemOpen
  \bibfield  {author} {\bibinfo {author} {\bibfnamefont {R.~P.}\ \bibnamefont
  {Elliot}},\ }\href@noop {} {\emph {\bibinfo {title} {Constitution of Binary
  Alloys, Suppl 1}}}\ (\bibinfo  {publisher} {McGraw-Hill: New York},\ \bibinfo
  {address} {San Diego, CA},\ \bibinfo {year} {1965})\BibitemShut {NoStop}%
\bibitem [{\citenamefont {Herfort}\ \emph {et~al.}(2004)\citenamefont
  {Herfort}, \citenamefont {Schoenherr}, \citenamefont {Friedland},\ and\
  \citenamefont {Ploog}}]{Herfort2004}%
  \BibitemOpen
  \bibfield  {author} {\bibinfo {author} {\bibfnamefont {J.}~\bibnamefont
  {Herfort}}, \bibinfo {author} {\bibfnamefont {H.~P.}\ \bibnamefont
  {Schoenherr}}, \bibinfo {author} {\bibfnamefont {K.~J.}\ \bibnamefont
  {Friedland}}, \ and\ \bibinfo {author} {\bibfnamefont {K.~H.}\ \bibnamefont
  {Ploog}},\ }\href@noop {} {\bibfield  {journal} {\bibinfo  {journal} {J. Vac.
  Sci. Technol. B}\ }\textbf {\bibinfo {volume} {22}},\ \bibinfo {pages} {2073}
  (\bibinfo {year} {2004})}\BibitemShut {NoStop}%
\bibitem [{\citenamefont {Herfort}\ \emph {et~al.}(2003)\citenamefont
  {Herfort}, \citenamefont {Sch\"onherr},\ and\ \citenamefont
  {Ploog}}]{herfort03}%
  \BibitemOpen
  \bibfield  {author} {\bibinfo {author} {\bibfnamefont {J.}~\bibnamefont
  {Herfort}}, \bibinfo {author} {\bibfnamefont {H.-P.}\ \bibnamefont
  {Sch\"onherr}}, \ and\ \bibinfo {author} {\bibfnamefont {K.~H.}\ \bibnamefont
  {Ploog}},\ }\href@noop {} {\bibfield  {journal} {\bibinfo  {journal} {Appl.
  Phys. Lett.}\ }\textbf {\bibinfo {volume} {83}},\ \bibinfo {pages} {3912}
  (\bibinfo {year} {2003})}\BibitemShut {NoStop}%
\bibitem [{\citenamefont {Hilse}\ \emph {et~al.}(2013)\citenamefont {Hilse},
  \citenamefont {Herfort}, \citenamefont {Jenichen}, \citenamefont {Trampert},
  \citenamefont {Hanke}, \citenamefont {Schaaf}, \citenamefont {Geelhaar},\
  and\ \citenamefont {Riechert}}]{hilse2013}%
  \BibitemOpen
  \bibfield  {author} {\bibinfo {author} {\bibfnamefont {M.}~\bibnamefont
  {Hilse}}, \bibinfo {author} {\bibfnamefont {J.}~\bibnamefont {Herfort}},
  \bibinfo {author} {\bibfnamefont {B.}~\bibnamefont {Jenichen}}, \bibinfo
  {author} {\bibfnamefont {A.}~\bibnamefont {Trampert}}, \bibinfo {author}
  {\bibfnamefont {M.}~\bibnamefont {Hanke}}, \bibinfo {author} {\bibfnamefont
  {P.}~\bibnamefont {Schaaf}}, \bibinfo {author} {\bibfnamefont
  {L.}~\bibnamefont {Geelhaar}}, \ and\ \bibinfo {author} {\bibfnamefont
  {H.}~\bibnamefont {Riechert}},\ }\href@noop {} {\bibfield  {journal}
  {\bibinfo  {journal} {Nano Lett.}\ }\textbf {\bibinfo {volume} {13}},\
  \bibinfo {pages} {6203} (\bibinfo {year} {2013})}\BibitemShut {NoStop}%
\bibitem [{\citenamefont {Wagner}\ and\ \citenamefont
  {Ellis}(1964)}]{Wagner1964}%
  \BibitemOpen
  \bibfield  {author} {\bibinfo {author} {\bibfnamefont {R.~S.}\ \bibnamefont
  {Wagner}}\ and\ \bibinfo {author} {\bibfnamefont {W.~S.}\ \bibnamefont
  {Ellis}},\ }\href@noop {} {\bibfield  {journal} {\bibinfo  {journal} {Appl.
  Phys. Lett.}\ }\textbf {\bibinfo {volume} {4}},\ \bibinfo {pages} {89}
  (\bibinfo {year} {1964})}\BibitemShut {NoStop}%
\bibitem [{\citenamefont {Mandl}\ \emph {et~al.}(2006)\citenamefont {Mandl},
  \citenamefont {Stangl}, \citenamefont {Martensson}, \citenamefont
  {Mikkelsen}, \citenamefont {Eriksson}, \citenamefont {Karlsson},
  \citenamefont {Bauer}, \citenamefont {Samuelson},\ and\ \citenamefont
  {Seifert.}}]{Mandl2006}%
  \BibitemOpen
  \bibfield  {author} {\bibinfo {author} {\bibfnamefont {B.}~\bibnamefont
  {Mandl}}, \bibinfo {author} {\bibfnamefont {J.}~\bibnamefont {Stangl}},
  \bibinfo {author} {\bibfnamefont {T.}~\bibnamefont {Martensson}}, \bibinfo
  {author} {\bibfnamefont {A.}~\bibnamefont {Mikkelsen}}, \bibinfo {author}
  {\bibfnamefont {J.}~\bibnamefont {Eriksson}}, \bibinfo {author}
  {\bibfnamefont {L.~S.}\ \bibnamefont {Karlsson}}, \bibinfo {author}
  {\bibfnamefont {G.}~\bibnamefont {Bauer}}, \bibinfo {author} {\bibfnamefont
  {L.}~\bibnamefont {Samuelson}}, \ and\ \bibinfo {author} {\bibfnamefont
  {W.}~\bibnamefont {Seifert.}},\ }\href@noop {} {\bibfield  {journal}
  {\bibinfo  {journal} {Nano Lett.}\ }\textbf {\bibinfo {volume} {6}},\
  \bibinfo {pages} {1817} (\bibinfo {year} {2006})}\BibitemShut {NoStop}%
\bibitem [{\citenamefont {Glas}\ \emph {et~al.}(2007)\citenamefont {Glas},
  \citenamefont {Harmand},\ and\ \citenamefont {Patriarche}}]{glas2007}%
  \BibitemOpen
  \bibfield  {author} {\bibinfo {author} {\bibfnamefont {F.}~\bibnamefont
  {Glas}}, \bibinfo {author} {\bibfnamefont {J.~C.}\ \bibnamefont {Harmand}}, \
  and\ \bibinfo {author} {\bibfnamefont {G.}~\bibnamefont {Patriarche}},\
  }\href@noop {} {\bibfield  {journal} {\bibinfo  {journal} {Phys. Rev. Lett.}\
  }\textbf {\bibinfo {volume} {99}},\ \bibinfo {pages} {146101} (\bibinfo
  {year} {2007})}\BibitemShut {NoStop}%
\bibitem [{\citenamefont {Fontcuberta}\ \emph {et~al.}(2008)\citenamefont
  {Fontcuberta}, \citenamefont {Colombo}, \citenamefont {Abstreiter},
  \citenamefont {Arbiol},\ and\ \citenamefont {Morante}}]{Fontcuberta2008}%
  \BibitemOpen
  \bibfield  {author} {\bibinfo {author} {\bibfnamefont {A.}~\bibnamefont
  {Fontcuberta}}, \bibinfo {author} {\bibfnamefont {C.}~\bibnamefont
  {Colombo}}, \bibinfo {author} {\bibfnamefont {G.}~\bibnamefont {Abstreiter}},
  \bibinfo {author} {\bibfnamefont {J.}~\bibnamefont {Arbiol}}, \ and\ \bibinfo
  {author} {\bibfnamefont {J.~R.}\ \bibnamefont {Morante}},\ }\href@noop {}
  {\bibfield  {journal} {\bibinfo  {journal} {Appl. Phys. Lett.}\ }\textbf
  {\bibinfo {volume} {92}},\ \bibinfo {pages} {063112} (\bibinfo {year}
  {2008})}\BibitemShut {NoStop}%
\bibitem [{\citenamefont {Wacaser}\ \emph {et~al.}(2009)\citenamefont
  {Wacaser}, \citenamefont {Dick}, \citenamefont {Johansson}, \citenamefont
  {Borgstroem}, \citenamefont {Deppert},\ and\ \citenamefont
  {Samuelson}}]{Wacaser2009}%
  \BibitemOpen
  \bibfield  {author} {\bibinfo {author} {\bibfnamefont {B.~A.}\ \bibnamefont
  {Wacaser}}, \bibinfo {author} {\bibfnamefont {K.~A.}\ \bibnamefont {Dick}},
  \bibinfo {author} {\bibfnamefont {J.}~\bibnamefont {Johansson}}, \bibinfo
  {author} {\bibfnamefont {M.~T.}\ \bibnamefont {Borgstroem}}, \bibinfo
  {author} {\bibfnamefont {K.}~\bibnamefont {Deppert}}, \ and\ \bibinfo
  {author} {\bibfnamefont {L.}~\bibnamefont {Samuelson}},\ }\href@noop {}
  {\bibfield  {journal} {\bibinfo  {journal} {Advanced Materials}\ }\textbf
  {\bibinfo {volume} {21}},\ \bibinfo {pages} {153} (\bibinfo {year}
  {2009})}\BibitemShut {NoStop}%
\bibitem [{\citenamefont {Takemoto}\ \emph {et~al.}(2006)\citenamefont
  {Takemoto}, \citenamefont {Murakami}, \citenamefont {Iwamoto}, \citenamefont
  {Matsuo}, \citenamefont {Kangawa}, \citenamefont {Kumagai},\ and\
  \citenamefont {Koukitu}}]{takemoto2006}%
  \BibitemOpen
  \bibfield  {author} {\bibinfo {author} {\bibfnamefont {K.}~\bibnamefont
  {Takemoto}}, \bibinfo {author} {\bibfnamefont {H.}~\bibnamefont {Murakami}},
  \bibinfo {author} {\bibfnamefont {T.}~\bibnamefont {Iwamoto}}, \bibinfo
  {author} {\bibfnamefont {Y.}~\bibnamefont {Matsuo}}, \bibinfo {author}
  {\bibfnamefont {Y.}~\bibnamefont {Kangawa}}, \bibinfo {author} {\bibfnamefont
  {Y.}~\bibnamefont {Kumagai}}, \ and\ \bibinfo {author} {\bibfnamefont
  {A.}~\bibnamefont {Koukitu}},\ }\href@noop {} {\bibfield  {journal} {\bibinfo
   {journal} {Jap. J. Appl. Phys.}\ }\textbf {\bibinfo {volume} {45}},\
  \bibinfo {pages} {478} (\bibinfo {year} {2006})}\BibitemShut {NoStop}%
\bibitem [{\citenamefont {vanderZiel}\ \emph {et~al.}(1987)\citenamefont
  {vanderZiel}, \citenamefont {Dupuis}, \citenamefont {Logan},\ and\
  \citenamefont {Pinzone}}]{Pinzone1987}%
  \BibitemOpen
  \bibfield  {author} {\bibinfo {author} {\bibfnamefont {J.~P.}\ \bibnamefont
  {vanderZiel}}, \bibinfo {author} {\bibfnamefont {R.~D.}\ \bibnamefont
  {Dupuis}}, \bibinfo {author} {\bibfnamefont {R.~A.}\ \bibnamefont {Logan}}, \
  and\ \bibinfo {author} {\bibfnamefont {J.}~\bibnamefont {Pinzone}},\
  }\href@noop {} {\bibfield  {journal} {\bibinfo  {journal} {Appl. Phys.
  Lett.}\ }\textbf {\bibinfo {volume} {51}},\ \bibinfo {pages} {89} (\bibinfo
  {year} {1987})}\BibitemShut {NoStop}%
\bibitem [{\citenamefont {Sakai}\ \emph {et~al.}(1988)\citenamefont {Sakai},
  \citenamefont {Matyi},\ and\ \citenamefont {Shichijo}}]{Sakai1988}%
  \BibitemOpen
  \bibfield  {author} {\bibinfo {author} {\bibfnamefont {S.}~\bibnamefont
  {Sakai}}, \bibinfo {author} {\bibfnamefont {R.~J.}\ \bibnamefont {Matyi}}, \
  and\ \bibinfo {author} {\bibfnamefont {H.}~\bibnamefont {Shichijo}},\
  }\href@noop {} {\bibfield  {journal} {\bibinfo  {journal} {Appl. Phys.
  Lett.}\ }\textbf {\bibinfo {volume} {63}},\ \bibinfo {pages} {1075} (\bibinfo
  {year} {1988})}\BibitemShut {NoStop}%
\bibitem [{\citenamefont {Zhu}\ \emph {et~al.}(2013)\citenamefont {Zhu},
  \citenamefont {Wallis},\ and\ \citenamefont {Humphreys}}]{Zhu2013}%
  \BibitemOpen
  \bibfield  {author} {\bibinfo {author} {\bibfnamefont {D.}~\bibnamefont
  {Zhu}}, \bibinfo {author} {\bibfnamefont {D.~J.}\ \bibnamefont {Wallis}}, \
  and\ \bibinfo {author} {\bibfnamefont {C.~J.}\ \bibnamefont {Humphreys}},\
  }\href@noop {} {\bibfield  {journal} {\bibinfo  {journal} {Rep. Prog. Phys.}\
  }\textbf {\bibinfo {volume} {76}},\ \bibinfo {pages} {106501} (\bibinfo
  {year} {2013})}\BibitemShut {NoStop}%
\bibitem [{\citenamefont {Kaganer}\ \emph {et~al.}(2009)\citenamefont
  {Kaganer}, \citenamefont {Jenichen}, \citenamefont {Shayduk}, \citenamefont
  {Braun},\ and\ \citenamefont {Riechert}}]{kag09}%
  \BibitemOpen
  \bibfield  {author} {\bibinfo {author} {\bibfnamefont {V.~M.}\ \bibnamefont
  {Kaganer}}, \bibinfo {author} {\bibfnamefont {B.}~\bibnamefont {Jenichen}},
  \bibinfo {author} {\bibfnamefont {R.}~\bibnamefont {Shayduk}}, \bibinfo
  {author} {\bibfnamefont {W.}~\bibnamefont {Braun}}, \ and\ \bibinfo {author}
  {\bibfnamefont {H.}~\bibnamefont {Riechert}},\ }\href@noop {} {\bibfield
  {journal} {\bibinfo  {journal} {Phys. Rev. Lett.}\ }\textbf {\bibinfo
  {volume} {102}},\ \bibinfo {pages} {016103} (\bibinfo {year}
  {2009})}\BibitemShut {NoStop}%
\bibitem [{\citenamefont {Copel}\ \emph {et~al.}(1989)\citenamefont {Copel},
  \citenamefont {Reuter}, \citenamefont {Kaxiras},\ and\ \citenamefont
  {Tromp}}]{Copel1989}%
  \BibitemOpen
  \bibfield  {author} {\bibinfo {author} {\bibfnamefont {M.}~\bibnamefont
  {Copel}}, \bibinfo {author} {\bibfnamefont {M.~C.}\ \bibnamefont {Reuter}},
  \bibinfo {author} {\bibfnamefont {E.}~\bibnamefont {Kaxiras}}, \ and\
  \bibinfo {author} {\bibfnamefont {R.~M.}\ \bibnamefont {Tromp}},\ }\href@noop
  {} {\bibfield  {journal} {\bibinfo  {journal} {Phys. Rev. Lett.}\ }\textbf
  {\bibinfo {volume} {63}},\ \bibinfo {pages} {632} (\bibinfo {year}
  {1989})}\BibitemShut {NoStop}%
\bibitem [{\citenamefont {LeGoues}\ \emph {et~al.}(1990)\citenamefont
  {LeGoues}, \citenamefont {Copel},\ and\ \citenamefont {Tromp}}]{LeGoues1990}%
  \BibitemOpen
  \bibfield  {author} {\bibinfo {author} {\bibfnamefont {F.~K.}\ \bibnamefont
  {LeGoues}}, \bibinfo {author} {\bibfnamefont {M.}~\bibnamefont {Copel}}, \
  and\ \bibinfo {author} {\bibfnamefont {R.~M.}\ \bibnamefont {Tromp}},\
  }\href@noop {} {\bibfield  {journal} {\bibinfo  {journal} {Phys. Rev. B}\
  }\textbf {\bibinfo {volume} {42}},\ \bibinfo {pages} {11690} (\bibinfo {year}
  {1990})}\BibitemShut {NoStop}%
\bibitem [{\citenamefont {von Hoegen}\ \emph {et~al.}(1991)\citenamefont {von
  Hoegen}, \citenamefont {LeGoues}, \citenamefont {Copel}, \citenamefont
  {Reuter},\ and\ \citenamefont {Tromp}}]{hoegen1991}%
  \BibitemOpen
  \bibfield  {author} {\bibinfo {author} {\bibfnamefont {M.~H.}\ \bibnamefont
  {von Hoegen}}, \bibinfo {author} {\bibfnamefont {F.~K.}\ \bibnamefont
  {LeGoues}}, \bibinfo {author} {\bibfnamefont {M.}~\bibnamefont {Copel}},
  \bibinfo {author} {\bibfnamefont {M.~C.}\ \bibnamefont {Reuter}}, \ and\
  \bibinfo {author} {\bibfnamefont {R.~M.}\ \bibnamefont {Tromp}},\ }\href@noop
  {} {\bibfield  {journal} {\bibinfo  {journal} {Phys. Rev. Lett.}\ }\textbf
  {\bibinfo {volume} {67}},\ \bibinfo {pages} {1130} (\bibinfo {year}
  {1991})}\BibitemShut {NoStop}%
\bibitem [{\citenamefont {Tromp}\ and\ \citenamefont
  {Reuter}(1992)}]{Tromp1992}%
  \BibitemOpen
  \bibfield  {author} {\bibinfo {author} {\bibfnamefont {R.~M.}\ \bibnamefont
  {Tromp}}\ and\ \bibinfo {author} {\bibfnamefont {M.~C.}\ \bibnamefont
  {Reuter}},\ }\href@noop {} {\bibfield  {journal} {\bibinfo  {journal} {Phys.
  Rev. Lett.}\ }\textbf {\bibinfo {volume} {68}},\ \bibinfo {pages} {954}
  (\bibinfo {year} {1992})}\BibitemShut {NoStop}%
\bibitem [{\citenamefont {van~der Vegt}\ \emph {et~al.}(1992)\citenamefont
  {van~der Vegt}, \citenamefont {van Pinxteren}, \citenamefont {Lohmeier},
  \citenamefont {Vlieg},\ and\ \citenamefont {Thornton}}]{vegt1992}%
  \BibitemOpen
  \bibfield  {author} {\bibinfo {author} {\bibfnamefont {H.~A.}\ \bibnamefont
  {van~der Vegt}}, \bibinfo {author} {\bibfnamefont {H.~M.}\ \bibnamefont {van
  Pinxteren}}, \bibinfo {author} {\bibfnamefont {M.}~\bibnamefont {Lohmeier}},
  \bibinfo {author} {\bibfnamefont {E.}~\bibnamefont {Vlieg}}, \ and\ \bibinfo
  {author} {\bibfnamefont {J.~M.~C.}\ \bibnamefont {Thornton}},\ }\href@noop {}
  {\bibfield  {journal} {\bibinfo  {journal} {Phys. Rev. Lett.}\ }\textbf
  {\bibinfo {volume} {68}},\ \bibinfo {pages} {3335} (\bibinfo {year}
  {1992})}\BibitemShut {NoStop}%
\bibitem [{\citenamefont {Kageyama}\ \emph {et~al.}(2004)\citenamefont
  {Kageyama}, \citenamefont {Miyamoto}, \citenamefont {Ohta}, \citenamefont
  {Matsuura}, \citenamefont {Matsui}, \citenamefont {Furuhata},\ and\
  \citenamefont {Koyama}}]{Kageyama2004}%
  \BibitemOpen
  \bibfield  {author} {\bibinfo {author} {\bibfnamefont {T.}~\bibnamefont
  {Kageyama}}, \bibinfo {author} {\bibfnamefont {T.}~\bibnamefont {Miyamoto}},
  \bibinfo {author} {\bibfnamefont {M.}~\bibnamefont {Ohta}}, \bibinfo {author}
  {\bibfnamefont {T.}~\bibnamefont {Matsuura}}, \bibinfo {author}
  {\bibfnamefont {Y.}~\bibnamefont {Matsui}}, \bibinfo {author} {\bibfnamefont
  {T.}~\bibnamefont {Furuhata}}, \ and\ \bibinfo {author} {\bibfnamefont
  {F.}~\bibnamefont {Koyama}},\ }\href@noop {} {\bibfield  {journal} {\bibinfo
  {journal} {J. Appl. Phys.}\ }\textbf {\bibinfo {volume} {96}},\ \bibinfo
  {pages} {44} (\bibinfo {year} {2004})}\BibitemShut {NoStop}%
\bibitem [{\citenamefont {Tixier}\ \emph {et~al.}(2003)\citenamefont {Tixier},
  \citenamefont {Adamcyk}, \citenamefont {Youngb}, \citenamefont {Schmid},\
  and\ \citenamefont {Tiedje}}]{Tixier2003}%
  \BibitemOpen
  \bibfield  {author} {\bibinfo {author} {\bibfnamefont {S.}~\bibnamefont
  {Tixier}}, \bibinfo {author} {\bibfnamefont {M.}~\bibnamefont {Adamcyk}},
  \bibinfo {author} {\bibfnamefont {E.}~\bibnamefont {Youngb}}, \bibinfo
  {author} {\bibfnamefont {J.}~\bibnamefont {Schmid}}, \ and\ \bibinfo {author}
  {\bibfnamefont {T.}~\bibnamefont {Tiedje}},\ }\href@noop {} {\bibfield
  {journal} {\bibinfo  {journal} {J. Cryst. Gr.}\ }\textbf {\bibinfo {volume}
  {251}},\ \bibinfo {pages} {449} (\bibinfo {year} {2003})}\BibitemShut
  {NoStop}%
\bibitem [{\citenamefont {Grandjean}\ \emph {et~al.}(1992)\citenamefont
  {Grandjean}, \citenamefont {Massies},\ and\ \citenamefont
  {Etgens}}]{Grandjean1992}%
  \BibitemOpen
  \bibfield  {author} {\bibinfo {author} {\bibfnamefont {N.}~\bibnamefont
  {Grandjean}}, \bibinfo {author} {\bibfnamefont {J.}~\bibnamefont {Massies}},
  \ and\ \bibinfo {author} {\bibfnamefont {V.~H.}\ \bibnamefont {Etgens}},\
  }\href@noop {} {\bibfield  {journal} {\bibinfo  {journal} {Phys. Rev. Lett.}\
  }\textbf {\bibinfo {volume} {69}},\ \bibinfo {pages} {796} (\bibinfo {year}
  {1992})}\BibitemShut {NoStop}%
\bibitem [{\citenamefont {Massies}\ and\ \citenamefont
  {Grandjean}(1993)}]{Massies1993}%
  \BibitemOpen
  \bibfield  {author} {\bibinfo {author} {\bibfnamefont {J.}~\bibnamefont
  {Massies}}\ and\ \bibinfo {author} {\bibfnamefont {N.}~\bibnamefont
  {Grandjean}},\ }\href@noop {} {\bibfield  {journal} {\bibinfo  {journal}
  {Phys. Rev. B}\ }\textbf {\bibinfo {volume} {48}},\ \bibinfo {pages} {8502}
  (\bibinfo {year} {1993})}\BibitemShut {NoStop}%
\bibitem [{\citenamefont {Teichert}\ \emph {et~al.}(2002)\citenamefont
  {Teichert}, \citenamefont {Hortenbach}, \citenamefont {Beddies},\ and\
  \citenamefont {Hinneberg}}]{Teichert2002}%
  \BibitemOpen
  \bibfield  {author} {\bibinfo {author} {\bibfnamefont {S.}~\bibnamefont
  {Teichert}}, \bibinfo {author} {\bibfnamefont {H.}~\bibnamefont
  {Hortenbach}}, \bibinfo {author} {\bibfnamefont {G.}~\bibnamefont {Beddies}},
  \ and\ \bibinfo {author} {\bibfnamefont {H.-J.}\ \bibnamefont {Hinneberg}},\
  }\href@noop {} {\bibfield  {journal} {\bibinfo  {journal} {Microelectronic
  Engineering}\ }\textbf {\bibinfo {volume} {60}},\ \bibinfo {pages} {255}
  (\bibinfo {year} {2002})}\BibitemShut {NoStop}%
\end{thebibliography}
%merlin.mbs apsrev4-1.bst 2010-07-25 4.21a (PWD, AO, DPC) hacked
%Control: key (0)
%Control: author (8) initials jnrlst
%Control: editor formatted (1) identically to author
%Control: production of article title (-1) disabled
%Control: page (0) single
%Control: year (1) truncated
%Control: production of eprint (0) enabled
%

%%\newpage

\end{document}